\documentclass[preprint,showpacs,preprintnumbers,amsmath,amssymb]{revtex4}
\usepackage{graphicx}
\usepackage{amsmath}
\usepackage{amssymb}
\usepackage{setspace}  
\usepackage{natbib}  
\usepackage{color}
\begin{document}
\newcommand{\blue}[1]{\textcolor{blue}{#1}}
\newcommand{\green}[1]{\textcolor{green}{#1}}
\newcommand{\red}[1]{\textcolor{red}{#1}}
\definecolor{violet}{rgb}{0.55,0.14,1.0}
\newcommand{\violet}[1]{\textcolor{violet}{#1}}

\title{Rheology, Structure and Dynamics of Colloid-Polymer Mixtures: from Liquids to Gels}
\author{M. Laurati$^1$, G. Petekidis$^2$, N. Koumakis$^2$, F. Cardinaux$^1$, A. B. Schofield$^3$, J. M. Brader$^4$, M. Fuchs$^4$ and S. U. Egelhaaf$^1$}
\affiliation{$^1$ Condensed Matter Physics Laboratory, Heinrich-Heine University, Universit{\"a}tsstr. 1, 40225 D{\"u}sseldorf, Germany.\\
$^2$ IESL-FORTH and Department of Materials Science and Technology, University of Crete, Heraklion, Greece.\\ $^3$ SUPA, School of Physics, The University of Edinburgh, Mayfield Road, Edinburgh, EH9 3JZ, United Kingdom.\\ $^4$ Soft Matter Theory Group, University Konstanz, Konstanz, Germany.}

\begin{abstract}
We investigated the viscoelastic properties of colloid-polymer mixtures at intermediate colloid volume fraction and varying polymer concentrations, thereby tuning the attractive interactions. Within the examined range of polymer concentrations, the samples ranged from fluids to gels. Already in the liquid phase the viscoelastic properties significantly changed when approaching the gelation boundary, indicating the formation of clusters and transient networks. This is supported by an increasing correlation length of the density fluctuations, observed by static light scattering and microscopy. At the same time, the correlation function determined by dynamic light scattering completely decays, indicating the absence of dynamical arrest. Upon increasing the polymer concentration beyond the gelation boundary, the rheological properties changed qualitatively again, now they are consistent with the formation of colloidal gels. Our experimental results, namely the location of the gelation boundary as well as the elastic (storage) and viscous (loss) moduli, are compared to different theoretical models. These include consideration of the escape time as well as predictions for the viscoelastic moduli based on scaling relations and Mode Coupling Theories (MCT).
\end{abstract}

\pacs{62.20.-x,62.10.+s,64.70.pv}

\maketitle

\section{Introduction}
The mechanical properties of solids and liquids are very different. A solid responds elastically to a small deformation, while a liquid flows. Complex fluids, such as colloidal suspensions, polymers or surfactant solutions, have mechanical properties between those of elastic solids and viscous liquids, they are viscoelastic. Furthermore, they can be significantly perturbed by even modest mechanical forces. This causes a wealth of fascinating effects \cite{larson99}, but also provides a challenge to fundamental and applied research to understand their behavior under deformation and flow, i.e. their rheology. A detailed knowledge of their properties is crucial for many applications; complex fluids are extensively used in industrial products and processes \cite{larson99,coussot05}.

Among complex fluids, colloidal suspensions are frequently used as models of atomic systems whose interparticle interactions can be tuned \cite{pusey:leshouches,poon04}. These model systems allow the investigation of various fundamental phenomena, such as the equilibrium thermodynamics of gas, liquid and crystal phases and also the non-equilibrium behavior of gels and glasses. The latter are disordered solids which are dynamically arrested and long-lived. Although they can be formed at any colloid volume fraction \cite{poon02,shah_jcp02}, most studies have focused on either very large or small volume fractions.

At low colloid volume fractions, the interparticle attraction induces the formation of clusters \cite{stradner04,sedgwick04,sedgwick05} which may interconnect to create a space-spanning network \cite{krall98,lu06,emanuela08}. A connection between the gelation boundary and the spinodal line has been proposed \cite{dhont95,ilett95,poon97,verhaegh97,manley05} with spinodal decomposition driving cluster formation and gelation \cite{emanuela08}. The elasticity of gels is related to the connectivity of the network and the size of the clusters, i.e.~to the heterogeneous structure of the network \cite{krall98,trappe00,furst08,pantina07}.

In contrast, at large volume fractions amorphous solids are already formed in the absence of attraction due to crowding; repulsive hard sphere glasses \cite{pusey86,pusey87}. Increasing attraction strength initially causes melting of the glass, before for even larger attractions again amorphous solids, attractive glasses, form \cite{pham02,eckert02,dawson01}. Mode-coupling theory (MCT) has shown that the formation of repulsive and attractive glasses is caused by dynamical arrest due to caging and bonding, respectively \cite{vanmegen93,vanmegen94,pham02,dawson01,cates03}. These two arrest mechanisms seem to dominate also the mechanical response to deformations. The elastic properties are determined by the confinement of particles and can be rationalized in terms of the ratio between the energy and the volume characterizing the structural length of the system, which at high volume fractions coincides with the particle size \cite{mason95,bergenholtz03}. In addition, rheological and scattering experiments on repulsive \cite{petekidis02,petekidis04} and attractive glasses \cite{pham06,pham08} indicated one (in the former) or two (in the latter) step yielding related to cage and bond breaking.

In the region of intermediate colloid volume fractions the origin of the fluid-solid transition (even in the quiescent state) is still under debate and the rheological properties of the amorphous solid are hardly understood. It has been suggested by experiments and MCT calculations \cite{zukoski03,bergenholtz03,bergenholtz99} that the route to gelation is comparable to the one at higher volume fractions, i.e.~dynamical arrest. This is, however, in contradiction to recent simulations \cite{emanuela07,emanuela08} which predict an important role of spinodal decomposition also at intermediate volume fractions. Moreover, it is not clear whether elasticity is caused by similar mechanisms as in attractive glasses or by the connectivity of the network, as suggested by the structural heterogeneity observed in scattering \cite{shah_jcp03} and confocal microscopy experiments \cite{varadan03,smith07,dibble06}. Finally, the evolution of the elastic response inside the gel has been rarely studied. It has been recently suggested that the micromechanic response of gels with different interparticle attraction is strongly correlated with the 'clusterlike' or 'stringlike' nature of the gel structure \cite{lee08}.

Here we are interested in this intriguing region of intermediate colloid volume fraction. In particular, we investigate the transition from liquid-like to solid-like behavior upon increasing interparticle attraction. We use a mixture of nearly hard-sphere colloidal particles with non-adsorbing linear polymer \cite{poon02,anderson02,gast83,vincent88,ilett95}. The polymer induces a depletion attraction between the particles whose range and strength can be tuned by the polymer size and concentration, respectively. We investigate the static and dynamic properties of samples with an increasing polymer concentration, i.e.~increasing strength of the attractive interaction. Upon increasing polymer concentration, the system evolves from an equilibrium liquid to a non-equilibrium, dynamically arrested gel \cite{poon99,starrs02,poon02,shah_jcp02}. We use static and dynamic light scattering and microscopy to investigate this liquid-solid transition and determine structural parameters, such as the characteristic correlation length of density fluctuations, and dynamic properties, such as the collective dynamics. The samples are also subjected to shear and their mechanical response is determined by rheology. This combination allows us to relate the static and dynamic properties in the quiescent samples to the behavior under small deformation, i.e.~the linear rheological properties. We particularly focus on the evolution of the viscoelastic moduli as a function of increasing attraction, i.e.~on moving from liquid-like to solid-like samples and deeper into the gel region. Viscoelastic measurements approaching gelation are compared to different models: estimates of the escape time and scaling relations as well as MCT predictions for the time and polymer concentration dependence of the shear moduli. MCT has only recently been extended to describe dynamically arrested states under shear \cite{fuchs:jcp:05,fuchs:prl:02,fuchs:faraday:03,brader07,indrani95,miyazaki02,miyazaki06,zausch08,brader08}.

\section{Materials and Methods}

\subsection{\label{sec:samples}Samples}

We investigated mixtures of Polymethylmethacrylate (PMMA) colloids and linear Polystyrene (PS, from Polymer Laboratories)  dispersed in cis-decalin at a temperature $T=23^\circ$C. The hydrodynamic radius of the PMMA particles, $R = 137$~nm,  was determined by dynamic light scattering in the very dilute regime. The polydispersity of the colloids was not directly measured, but suppression of crystallization in quiescent and sheared colloidal dispersions indicates a polydispersity of about 12\%. The radius of gyration of the PS (molecular weight $M_{w} = 132.9$~kg/mol) in cis-decalin, $r_{g} = 10.8$ nm, was estimated based on \cite{berry66} and the polydispersity was cited as $M_w/M_n=1.01$. In dilute solution, this implies a polymer-colloid size ratio $\xi=r_{g}/R=0.079\pm 0.013$. The effective polymer-colloid size ratio $\xi^*$ takes into account the concentration dependence of the polymer size and the mesh size in the semidilute regime. It has been calculated according to the Generalized Free Volume Theory (GFVT) \cite{tuinier07,lekkerkerker92}.

The colloid stock solutions with volume fraction $\phi=0.6$ were prepared by redispersing spun-down sediments, whose volume fraction was estimated to be $\phi=0.67$ when taking polydispersity into account \cite{schaertl94}. Polymer stock solutions were prepared by adding cis-decalin to dry polymer. Polymer concentrations $c_{p}$ (mass/volume) were calculated from the weighed masses of the two components and their densities. Colloid-polymer mixtures were obtained by mixing appropriate amounts of colloid and polymer stock solutions. After mixing, samples were vigorously shaken using a vortex shaker, then homogenized over 3 days in a rotating wheel mixer. The compositions of the samples are summarized in table~\ref{tab_samples} with the first column giving the nominal polymer concentration which is used to refer to samples in the following. The polymer overlap concentration $c_{p}^{*}$ has been estimated by $c_{p}^{*}=3M_{w}/4\pi N_{A}r_{g}^{3}$. Values of $\phi$ and $c_{p}$ refer to the total volume and $c_{p}^{\mathrm{free}}$ to the volume not occupied by colloids as estimated by GFVT \cite{tuinier07,lekkerkerker92,aarts02}.

Samples were mixed and the light scattering or microscopy measurements perfomed within an hour to avoid effects due to aging (Sec.~\ref{sec:rheology}).

\begin{table}
\begin{center}
\begin{tabular}{| c | c c c c |}
\hline
$c_{p}/c_{p}^{*}$ nominal & $\phi$ & $c_{p}/c_{p}^{*}$ & $c_{p}^{\mathrm{free}}/c_{p}^{*}$ & $\xi^*$\\
\hline
0 & 0.40 & - & - & - \\
0.1 & 0.40 & 0.10 & 0.21 & 0.079 $\pm$ 0.01\\
0.2 & 0.40 & 0.20 & 0.40 & 0.067 $\pm$ 0.009\\
0.25 & 0.40 & 0.25 & 0.49 & 0.063 $\pm$ 0.008\\
0.32 & 0.39 & 0.32 & 0.62 & 0.057 $\pm$ 0.007\\
0.4 & 0.40 & 0.40 & 0.76 & 0.052 $\pm$ 0.007\\
0.5 & 0.41 & 0.48 & 0.90 & 0.048 $\pm$ 0.006\\
0.7 & 0.40 & 0.70 & 1.28 & 0.040 $\pm$ 0.006\\
0.8 & 0.40 & 0.82 & 1.48 & 0.037 $\pm$ 0.005\\
1 & 0.40 & 0.99 & 1.78 & 0.033 $\pm$ 0.005\\
1.5 & 0.40 & 1.49 & 2.63 & 0.027 $\pm$ 0.004\\
2 & 0.40 & 1.99 & 3.48 & 0.023 $\pm$ 0.004\\ 
\hline 
\end{tabular}
\caption{\label{tab_samples}List of samples. $\phi$ is the colloid volume fraction, $c_{p}/c_{p}^{*}$ and $c_{p}^{\mathrm{free}}/c_{p}^{*}$ are the polymer concentrations in the total and free volume, respectively, in units of the overlap concentration, $\xi^*$ is the effective polymer-colloid size ratio.}
\end{center}
\end{table}

\subsection{Light Scattering}
Due to the difference in refractive index $n$ between PMMA and cis-decalin ($n_{\mathrm{PMMA}}=1.49$, $n_{\mathrm{dec}}=1.48$) the samples are turbid. Multiple scattering was suppressed and single scattered light recorded using a 3D dynamic light scattering instrument (LS Instruments) \cite{schatzel91,urban98}.
From the cross-correlation functions we extracted the dynamic structure factors $f(Q,\tau)$ with the delay time $\tau$, the modulus of the scattering vector  $Q = (4n_{\mathrm{dec}}\pi/\lambda)\sin(\theta/2)$, the scattering angle $\theta$ and the wavelength $\lambda = 633$~nm (HeNe laser from JDS Uniphase).

In static light scattering experiments, the $Q$-dependence of the time-averaged intensity $\langle I(Q)\rangle$ was calculated from the time-averaged intensities recorded by the two detectors, $\langle I_a(Q)\rangle$ and $\langle I_b(Q)\rangle$, and the intercept $\beta_{ab}$ of the cross-correlation function:
\begin{equation}
\langle I(Q)\rangle = \sqrt{\langle I_a^{(1)}(Q)\rangle \langle I_b^{(1)}(Q)\rangle}=\sqrt{\langle I_a(Q)\rangle \langle I_b(Q)\rangle (\beta_{ab}/\beta_{ab}^{(1)})}
\label{eq3:statiq}
\end{equation}
where the superscript `(1)' refers to quantities determined in the single-scattering regime. To achieve ensemble averaging, the sample was rotated continuously. Rotation does not affect $\beta_{ab}$, $\langle I_a(Q)\rangle$ and $\langle I_b(Q)\rangle$, but the time dependence of $f(Q,\tau)$. Static structure factors $S(Q)$ were obtained from $\langle I(Q)\rangle$ taking into account the particle form factor (as determined in the dilute regime) and the transmissions $T_{a}$ and $T_{b}$ of the sample:
\begin{equation}
S(Q) = \frac {\phi^{\mathrm{(d)}}}{\phi} \frac{\sqrt{T_{a}^{\mathrm{(d)}} T_{b}^{\mathrm{(d)}}}}{\sqrt{T_{a}T_{b}}}\frac{I(Q)}{I^{\mathrm{(d)}}(Q)}
\label{eq4:statsq}
\end{equation}
where the superscript `(d)' refers to quantities determined in the dilute regime. This assumes that all significant contributions to the scattering are due to the colloids, as shown in \cite{pham04} for similar colloid-polymer mixtures (see also \cite{shah03,ballauff97}).

\subsection{Microscopy}

DIC microscopy experiments were perfomed using a Nikon Eclipse 80i upright microscope with a Nikon $100\times$ Plan Apo objective and a Canon EOS 30-D digital camera. Samples were loaded into a home-built cell: Two nr.~1 coverslips were glued onto a microscope slide, leaving a 3 to 4 mm wide channel between them. The channel was filled with the sample and a further nr.~1 coverslip used to cover the sample at the top and glue (UV-cure adhesive, Norland Optical Adhesive (NOA) 61) to seal the open ends of the channel \cite{jenkins08}.

\subsection{\label{sec:rheology}Rheology}

We used an ARES-HR rheometer with a force balance transducer 10FRTN1 and a cone-plate geometry (cone angle 0.044~rad, cone diameter 25~mm) which provides a constant strain throughout the sample. The geometry surfaces were mechanically roughened to avoid wall slip. To test reliability of the geometries with roughened surfaces, we compared results obtained with roughened and smooth surface geometries for samples where the presence of wall slip could be excluded. The agreement was found to be satisfactory. In dynamic measurements wall slip apparently has no effect in the linear viscoelastic regime, but dramatically affects measurements at large strains in the non-linear regime when the polymer concentration is comparable or larger than c$^*$. Here we only investigate the linear regime with strain amplitudes $0.001 \le \gamma_0 \le 0.02$, while the results in the non-linear regime are discussed elsewhere \cite{laurati08}.

In order to minimize solvent evaporation, a solvent saturation trap was used. The trap isolates the sample from the surrounding atmosphere by a fluid seal at the top and a permanent seal at the bottom. Solvent evaporation leads to a saturated atmosphere inside the enclosure.

In order to eliminate the effect of sample loading and aging, the following procedure was adopted: After loading, a dynamic strain sweep test was performed, i.e. the samples were subjected to oscillatory shear at a frequency of $\omega=1$ rad/s and the strain amplitude $\gamma_0$ was increased until the sample showed a liquid-like response; $\gamma_0 = 8$ was sufficient at all $c_p$. Moreover, before each test, oscillatory shear with $\omega=1$ rad/s and $\gamma_0 = 8$ was imposed on the samples until $G^{\prime}$ and $G^{\prime\prime}$ reached constant, steady-state values. Subsequently, samples were left at rest for a waiting time $t_w$ before the test was started. We performed aging experiments, a series of Dynamic Time Sweeps at $\omega=10$~rad/s and a total duration of 50000~s without any rejuvenation in between. Samples below the macroscopic gelation boundary showed no aging effects over the whole time interval, while gels showed an initial increase of the elastic modulus within the first 200~s after loading, but then the moduli remained constant at least up to 3600~s. A detailed study of aging effects on the rheological properties of gels will be reported elsewhere \cite{koumakis_aging}. Here we note that for $200\;{\mathrm{s}}\le t_w \le 3600$~s the viscoelastic properties did not change and reproducible results were obtained in consecutive tests. We have chosen $t_w=300$~s.

\section{Results and Discussion}

\subsection{Quiescent Samples}

\subsubsection{Macroscopic Behavior}

\begin{figure}[tbp]
\begin{center}
\includegraphics[scale=0.4,angle=270]{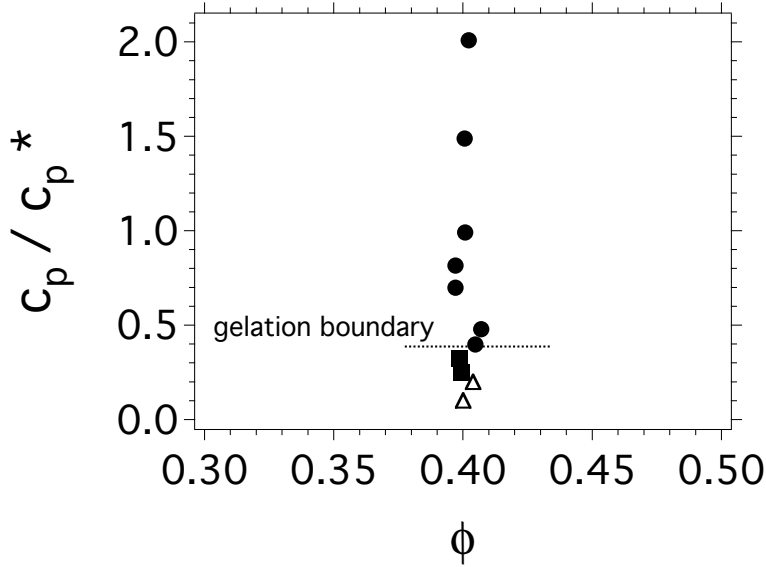}
\caption{Macroscopic  behaviour as investigated by tube inversion for different volume fractions $\phi$ and polymer concentrations $c_{p}/c_{p}^{*}$. ($\bullet$) gels (no flow), ($\blacksquare$) highly viscous fluids, ($\bigtriangleup$) low viscosity fluids.}
\label{fig3:macro_phase_diag}
\end{center}
\end{figure}
 
The macroscopic behavior of the samples was investigated by tube inversion as a function of colloid volume fraction $\phi$ and polymer concentration $c_{p}/c_{p}^{*}$ (Fig.~\ref{fig3:macro_phase_diag}). Gel samples were identified by the absence of flow after tube inversion, which was found for $c_{p}/c_{p}^{*}\ge 0.4$. Samples with $0.2<c_{p}/c_{p}^{*}<0.4$ showed already a relatively high viscosity, but were still flowing.

\subsubsection{\label{sec:sls}Microscopic Structure}

\begin{figure}[htbp!]
\begin{center}
\includegraphics[scale=0.45]{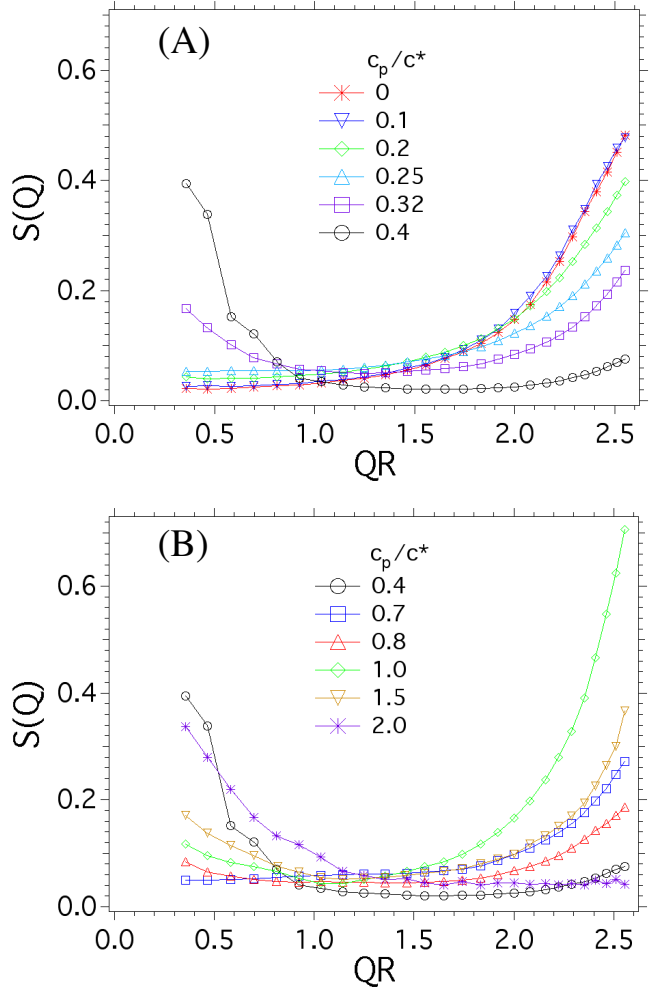}
\caption{Static structure factor $S(Q)$ measured by static light scattering for different polymer concentrations $c_{p}/c_{p}^{*}$ (as indicated in the legends) for samples below (a) and above (b) the macroscopic gelation boundary. }
\label{fig8:structure_factors}
\end{center}
\end{figure}

Microscopic structural information was obtained by static light scattering (SLS) and DIC microscopy. We determined the static structure factor $S(Q)$ at low scattering vectors $0.35\lesssim QR \lesssim 2.6$ (Fig.~\ref{fig8:structure_factors}) where length scales corresponding to collective structures of the order of a few particle diameters are probed. Due to the limited $Q$-range we cannot observe the first peak of $S(Q)$ which for a pure colloidal dispersion with $\phi = 0.4$ is expected, depending on polydispersity, in the range $3< QR <4$ \cite{shah03,dekruif88}.

Below the macroscopic gelation boundary ($c_{p}/c_{p}^{*}<0.4$, Fig.~\ref{fig8:structure_factors}A), $S(QR{<}1)$ monotonically increases with increasing $c_{p}/c_{p}^{*}$. For $c_{p}/c_{p}^{*}\le 0.25$ a finite value of S$(Q{\rightarrow}0)$ could be extrapolated which is consistent with the clustering of particles due to attractive depletion interactions, as has already been observed for silica-PS mixtures at the same colloid volume fraction \cite{shah03}. These clusters are not necessarily equilibrium clusters \cite{stradner04,sedgwick04,sedgwick05}. For larger polymer concentrations, $c_{p}/c_{p}^{*} = 0.32$ and $0.4$, $S(QR{<}1)$ increases steeply, which indicates an increasing amplitude of the density fluctuations. Crossing the gelation boundary ($c_{p}/c_{p}^{*}\approx 0.4$), $S(QR{<}1)$ drops dramatically pointing at the suppression of large density fluctuations. Then $S(QR{<}1)$ increases again inside the gel region, i.e.~for $c_{p}/c_{p}^{*}>0.4$ (Fig.~\ref{fig8:structure_factors}B).

\begin{figure}[tbp!]
\begin{center}
\includegraphics[scale=0.5]{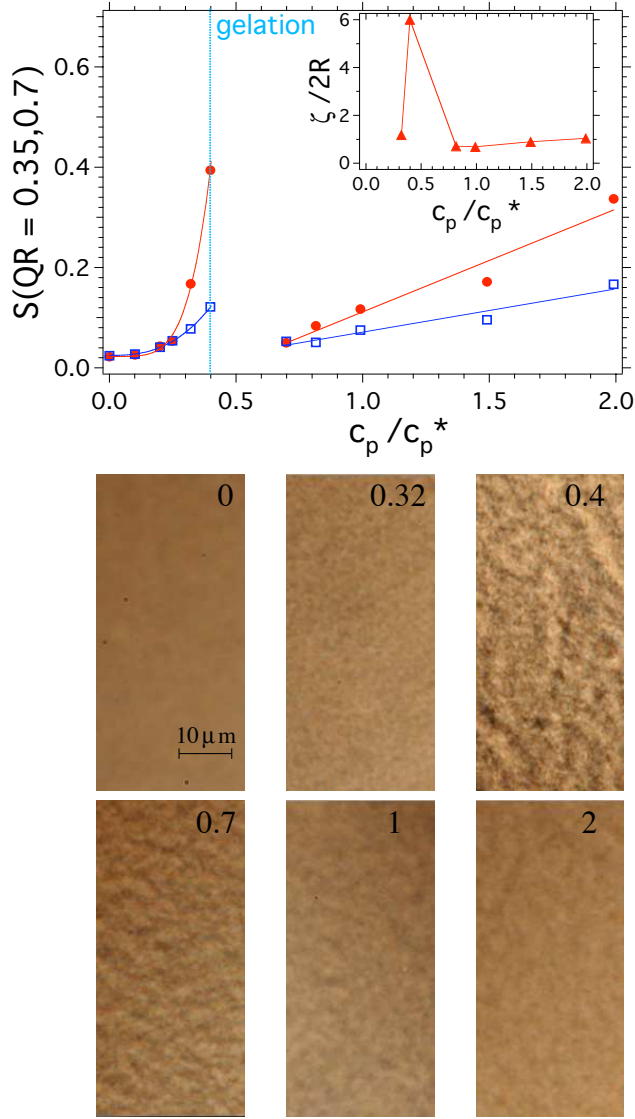}
\caption{Upper plot: Static structure factor $S(Q)$ at  $QR=0.35$ (red circles) and $QR=1.4$ (blue squares) as a function of polymer concentration $c_{p}/c_{p}^{*}$. Lines are fits by a power-law dependence for $c_{p}/c_{p}^{*} < 0.4$, and a linear dependence for $c_{p}/c_{p}^{*} > 0.4$. Inset: Correlation length $\zeta$ obtained by fitting an Ornstein-Zernike scaling to the low $Q$ part of $S(Q)$. Lower pictures: DIC microscopy images of samples with polymer concentrations $0\le c_{p}/c_{p}^{*}\le 2$ as indicated.}
\label{fig9:sq_lowq_zeta}
\end{center}
\end{figure}


The $c_{p}$-dependence of $S(Q)$ at two distinct $Q$ values ($QR=0.35$ and $0.7$) is summarized in Figure~\ref{fig9:sq_lowq_zeta}. The strong increase of $S(Q)$ upon approaching the gelation boundary, i.e.~$c_{p}/c_{p}^{*}\lesssim 0.4$, can be described by a power-law dependence, $S(Q) \sim (c_{p}/c_{p}^{*})^\alpha$ \cite{shah03}, with an exponent $\alpha_{0.35} = 4.6\pm 0.3$ for $QR=0.35$ and $\alpha_{0.7} = 2.6\pm 0.1$ for $QR=0.7$. After the sharp drop at the gelation boundary, $S(Q)$ increases roughly linearly with increasing $c_{p}/c_{p}^{*}$ inside the gel region. 

Fitting an Ornstein-Zernike scaling, $S(Q) \sim 1/[Q^2+(1/\zeta)^2]$ to those $S(QR{<}1)$ which increase at low $Q$, a characteristic correlation length $\zeta$ can be extracted (Fig.~\ref{fig9:sq_lowq_zeta}, inset). We find that $\zeta/2R$ increases from approximately 1 to 6 upon increasing $c_{p}/c_{p}^{*}$ from 0.32 to 0.4 and then, inside the gel region, drops again to approximately 1 with a slight increase with increasing $c_{p}/c_{p}^{*}$.

Due to the limited $Q$-range accessible in our light scattering experiments and, as a consequence, the large uncertainty in the value of $\zeta$, we complemented our light scattering experiments by DIC microscopy (Fig.~\ref{fig9:sq_lowq_zeta}). At $c_{p}/c_{p}^{*} = 0$ the sample appears homogeneous reflecting its fluid structure. Increasing $c_{p}/c_{p}^{*}$ toward the gel boundary, some graininess due to large scale structures is visible with the length scale and amplitude increasing strongly at the gelation boundary ($c_{p}/c_{p}^{*} = 0.4$). Within the gel phase ($c_{p}/c_{p}^{*} \ge 0.7$) the length scale and amplitude of the observed graininess decreases and subsequently saturates. DIC microscopy thus indicates that structural heterogeneities have a maximum around $c_{p}/c_{p}^{*} = 0.4$ and their length scale well before gelation and inside the gel region appears to be comparable. This is consistent with our light scattering results.

Large scale heterogeneous structures have been reported based on confocal microscopy experiments for gels composed of polymer-grafted silica spheres \cite{varadan03} and in dense PMMA-PS suspensions \cite{smith07,smith04} where the maximum degree of heterogeneity has also been observed in the vicinity of the gelation boundary. The sharp maximum of the structural correlation length at the gelation boundary is expected for an arrested phase separation \cite{emanuela07}. In this scenario, phase separation at the gelation boundary leads to a coarsening and cluster formation that is interrupted by dynamical arrest when the clusters permanently bond to form a gel. Due to our relatively large volume fraction, it is also conceivable that clusters connect to form a transient percolated network, which only arrests when the bond lifetime becomes large enough. To confirm that phase separation ultimately causes gelation is beyond the scope of this report and requires a more detailed study, including an investigation of the time dependence of the low $Q$ scattering after mixing.

\subsubsection{\label{sec:dls}Dynamics}

\begin{figure}[tbp]
\begin{center}
\includegraphics[scale=0.45,angle=270]{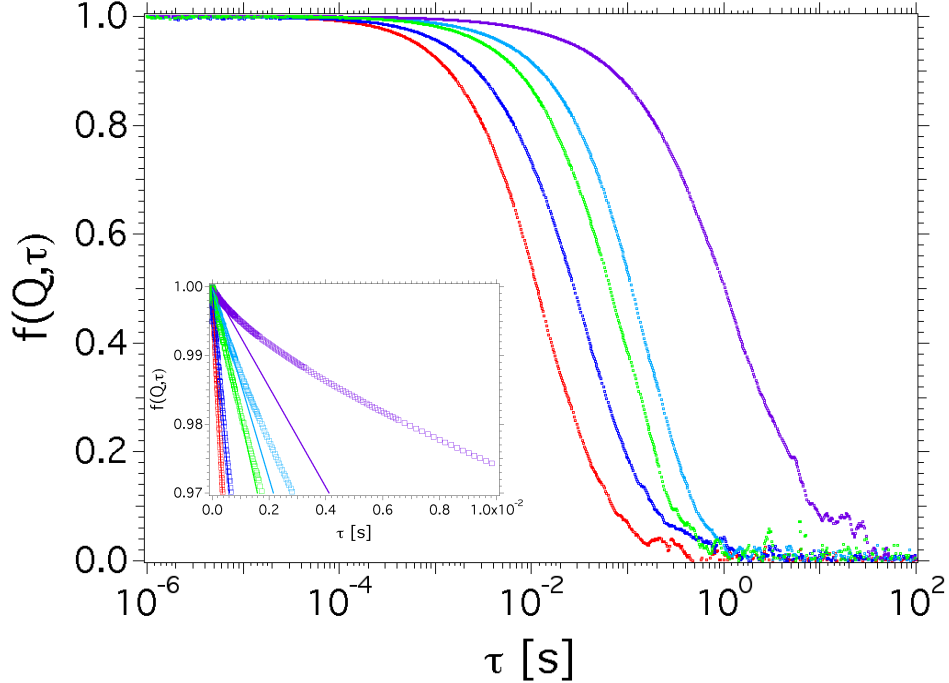}
\caption{ Dynamic structure factor $f(Q,\tau)$ at $QR=0.7$ as a function of delay time $\tau$ for samples with polymer concentration $c_{p}/c_{p}^{*}=0$ (red), 0.1 (blue), 0.2 (green), 0.25 (turquoise) and 0.32 (violet; from left to right). Inset:  Initial decay of $f(Q,\tau)$. Lines are fits to the short-time expansion of $f(Q,\tau)$ derived from the Smoluchowski equation \cite{pusey:leshouches}.}
\label{fig10:g2_dls}
\end{center}
\end{figure}

\begin{figure}[tbp]
\vspace{1cm}
\begin{center}
\includegraphics[scale=0.45,angle=270]{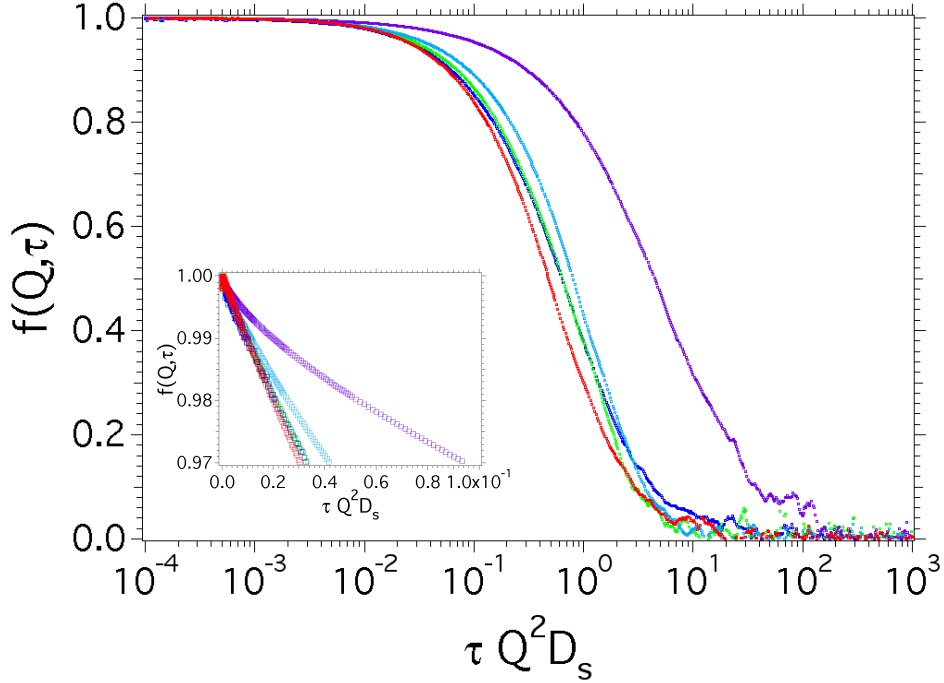}
\caption{Dynamic structure factor $f(Q,\tau)$ at $QR=0.7$ as a function of the rescaled delay time $\tau Q^2 D_S$ with the effective short-time diffusion coefficient $D_S$. Samples are as in figure \ref{fig10:g2_dls}. Inset:  Initial decay of $f(Q,\tau)$.}
\label{fig11:g2_dls_long}
\end{center}
\end{figure}

Dynamic light scattering (DLS) was performed with samples below the macroscopic gelation boundary ($c_{p}/c_{p}^{*} < 0.4$). Measurements were done at different scattering vectors $Q$, all of them below the first peak of the structure factor where the dynamic structure factor $f(Q,\tau)$ reflects collective dynamics. The data obtained at $QR = 0.7$ are shown in figures~\ref{fig10:g2_dls} and~\ref{fig11:g2_dls_long}. That $f(Q,\tau)$ completely decays indicates that the particle dynamics is ergodic, consistent with the macroscopic gelation boundary at $c_{p}/c_{p}^{*}>0.32$. The decay of $f(Q,\tau)$ cannot be described by a single exponential: At short times it is exponential, at long times stretched exponential. With increasing $c_{p}$ the particle dynamics is observed to slow down on both time scales (Fig.~\ref{fig10:g2_dls}). 

The initial, fast decay of $f(Q,\tau)$ covers only about 3\% of the total decay (Fig.~\ref{fig10:g2_dls}, inset). At short times, individual particles diffuse freely as reflected in the linear time dependence. This linear dependence is described by the short-time limit expression derived from the Smoluchowski equation \cite{pusey:leshouches}: $f(Q,\tau)=1-D_S(Q)Q^2\tau+O(\tau^2)$, where $D_S(Q)$ is the effective short-time diffusion coefficient. Departure from this free diffusion is observed at progressively shorter times for increasing $c_{p}$. This $c_{p}$-dependent departure is clearly visible if $f(Q,\tau)$ is plotted as a function of the rescaled time $\tau Q^2D_S$ (Fig.~\ref{fig11:g2_dls_long}, inset). The rescaling with $D_S$, whose Q-dependence is here omitted since $Q$ is fixed, also accounts for the trivial dependence on the viscosity of the polymer solution $\eta_r \eta_{\mathrm{dec}}$ ($\eta_{\mathrm{dec}}$ is the solvent viscosity).

The long-time relaxation is also slowed down (Figs.~\ref{fig10:g2_dls} and~\ref{fig11:g2_dls_long}). It corresponds to a collective, slow relaxation process related to the diffusion of particles whose movements are restricted by their mutual attraction. Its stretched form can be caused by size polydispersity, which leads to a spread in the long-time self diffusion coefficients \cite{pusey:leshouches}, and/or a distribution of particle diffusivities, caused by heterogeneities in the particle density, in agreement with the static light scattering data (Fig.~\ref{fig9:sq_lowq_zeta}).

\begin{figure}[tbp]
\begin{center}
\vspace{1cm}
\includegraphics[scale=0.55]{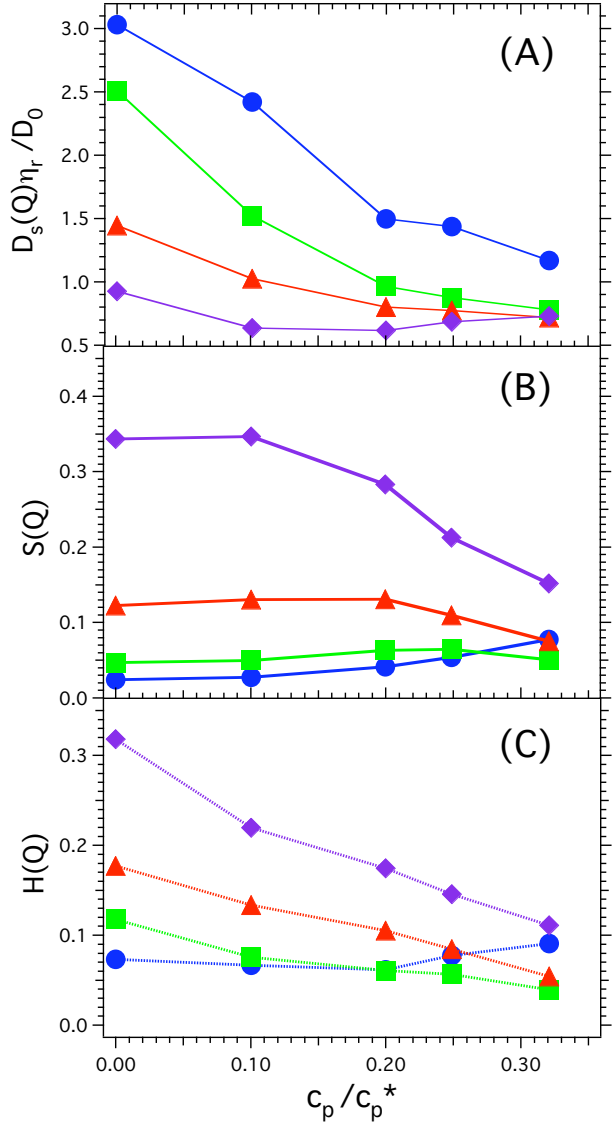}
\caption{(A) Normalized short-time diffusion coefficient $D_S(Q)\eta_r/D_0$, (B) structure factor $S(Q)$ and (C) hydrodynamic function $H(Q)$ as a function of polymer concentration $c_{p}/c_{p}^{*}$ for different scattering vectors $QR=0.7$ (\blue{$\bullet$}),  $1.35$ (\green{$\blacksquare$}), $1.9$ (\red{$\blacktriangle$}) and $2.35$ (\violet{$\blacklozenge$}).}
\label{fig12:g2_dls_param}
\end{center}
\end{figure}

\begin{figure}[tbp]
\begin{center}
\vspace{1cm}
\includegraphics[scale=0.55]{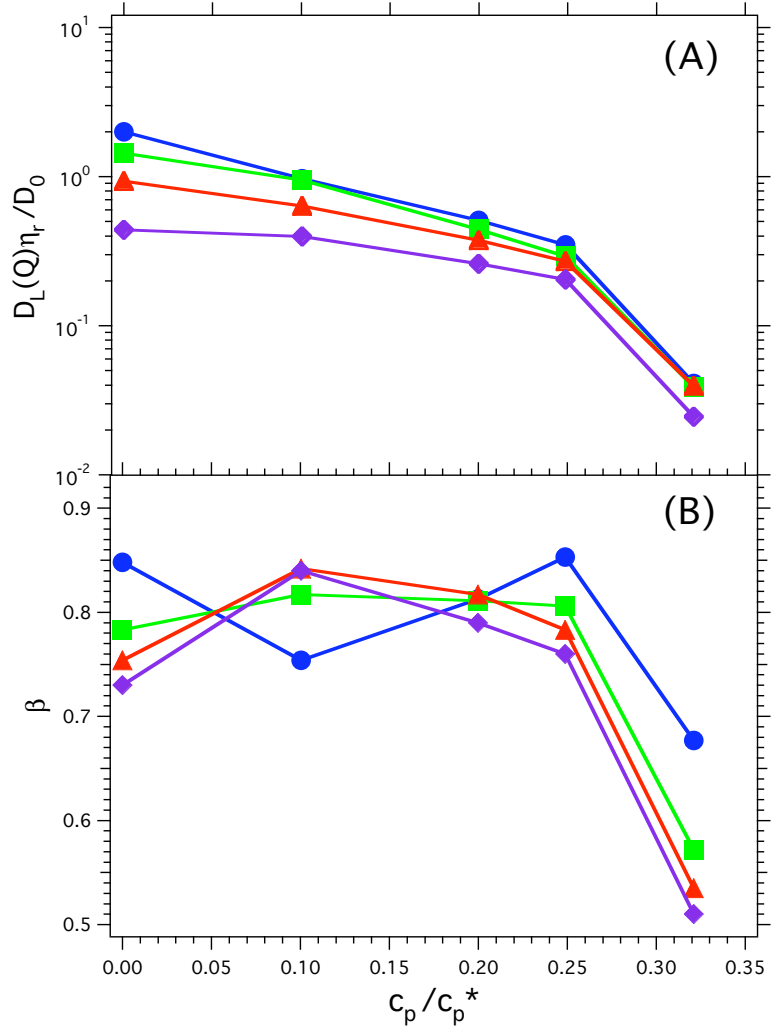}
\caption{(A) Normalized long-time diffusion coefficient $D_L(Q)\eta_r/D_0$ and (B) stretching exponents $\beta$ as a function of polymer concentration $c_p/c_p^*$ for different scattering vectors $QR=0.7$ (\blue{$\bullet$}),  $1.35$ ((\green{$\blacksquare$}), $1.9$ (\red{$\blacktriangle$}) and $2.35$ (\violet{$\blacklozenge$}).}
\label{fig14:g2_dls_param2}
\end{center}
\end{figure}

The short-time behavior for different $Q$ is summarized in figure \ref{fig12:g2_dls_param}. Shown is the $c_p$-dependence of the normalized short-time diffusion coefficient $D_S(Q)\eta_r/D_0$ where $D_0=\mathrm{k_B}T/6\pi\eta_{\mathrm{dec}}R$ is the free diffusion coefficient. It decreases with increasing $Q$ and $c_p$, especially in the range $0\le c_p/c_p^* \le 2$. These trends of $D_S(Q)\eta_r/D_0=H(Q)/S(Q)$ result from a delicate balance between the structure factor $S(Q)$ and the hydrodynamic function $H(Q)$ (Fig.~\ref{fig12:g2_dls_param}B, C). $S(Q)$ has been determined by static light scattering (Fig.~\ref{fig8:structure_factors}). It is almost constant for $c_p/c_p^*\lesssim 0.2$ and then decreases toward the gelation boundary, except for the smallest scattering vector, $QR=0.7$ (Fig.~\ref{fig12:g2_dls_param}B). Based on $D_S(Q)\eta_r/D_0$ and $S(Q)$, $H(Q)$ was calculated (Fig.~\ref{fig12:g2_dls_param}C). Except for the smallest $Q$, $QR=0.7$, $H(Q)$ decreases monotonically with increasing $c_p$.

The values of $D_S(Q)\eta_r/D_0$ for small $c_p/c_p^*$, i.e.~close to the pure hard-sphere case, are larger than 1 due to the dominant contribution of $H(Q)$ as compared to $S(Q)$, in agreement with previous data \cite{pusey:leshouches}. Increasing attraction within $c_p/c_p^* \lesssim 0.2$ hardly affects the structure, hence $S(Q)$ is about constant, while it causes a decrease of $H(Q)$ which for $c_p/c_p^* \approx 0.2$ becomes smaller than $S(Q)$ resulting in a decrease of $D_S(Q)\eta_r/D_0$ to about 1. For larger polymer concentrations, $c_p/c_p^* \gtrsim 0.2$, the particle attraction starts to affect the average structure with a reduction of $S(Q)$ near the peak (large $QR$) and an increase of $S(Q)$ at small $QR$ due to aggregation (Fig.~\ref{fig8:structure_factors}A). Thus the trends of $H(Q)$ and $S(Q)$ become similar and $D_S(Q)\eta_r/D_0$ tends to a constant value. Only for the largest polymer concentration, the value of $D_S(Q)\eta_r/D_0$ remains approximately constant for $QR > 0.7$. A similar trend has been observed in the liquid phase of a phase separating colloid-polymer mixture at comparable volume fractions \cite{voudouris08}.

The normalized long-time diffusion coefficient $D_L(Q)\eta_r/D_0$ and stretching exponent $\beta$ are obtained by fitting a stretched exponential to the long-time behavior of the dynamic structure factor (Fig.~\ref{fig14:g2_dls_param2}). $D_L(Q)\eta_r/D_0$ shows first a modest and then a sharp decrease with increasing $c_p$. This indicates that particle attraction slows down particle motions and, in the vicinity of the gelation boundary ($c_{p}/c_{p}^{*} \lesssim 0.4$), the particles start to be localized, signaling the approach of dynamical arrest. The stretching exponent $\beta$ is always below 1 with an approximately constant value of about 0.8 for $c_{p}/c_{p}^{*}<0.25$ and a sharp decrease at $c_{p}/c_{p}^{*} = 0.32$. This indicates that, upon approaching the gelation boundary, density fluctuations increase and broaden the distribution of effective long-time diffusion coefficients, consistent with the increasing correlation length $\zeta$ of structural heterogeneities observed by static light scattering and microscopy (Fig.~\ref{fig9:sq_lowq_zeta}).

\subsection{\label{sec:exp_rheo}Samples under Shear}

\begin{figure}[tbp]
\begin{center}
\vspace{1cm}
\includegraphics[scale=0.6]{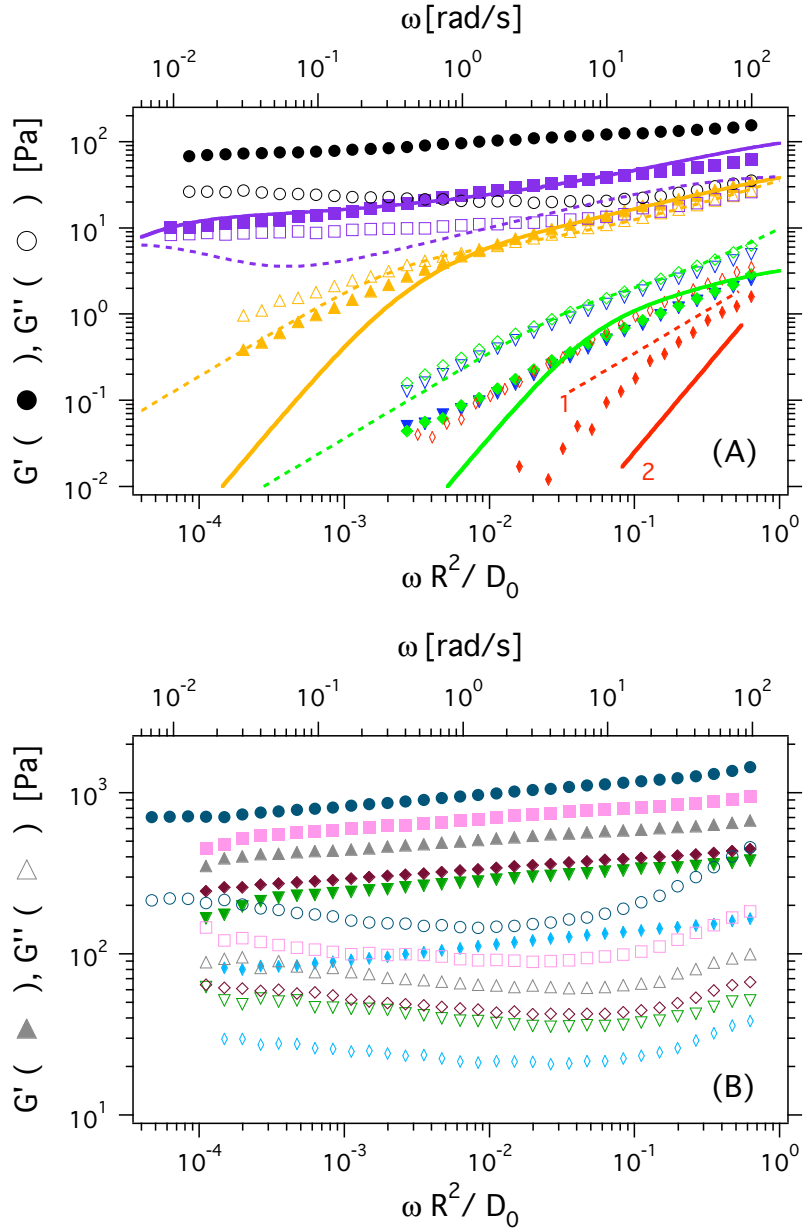}
\caption{\label{fig4:dfs_all_07}Dynamic frequency sweeps with elastic (storage) modulus $G^\prime$ (filled symbols) and viscous (loss) modulus $G^{\prime\prime}$ (open symbols) as a function of frequency $\omega$ in units of the instrument (top axis) and in units of the inverse diffusion time in the dilute limit $\tau_0=R^2/D_0\approx 4\times 10^{-2}$~s (bottom scale). Samples below the macroscopic gelation boundary ($c_p/c_p^*\le 0.4$) are shown in the upper, in the gel region ($c_p/c_p^* > 0.4$) in the lower plot with polymer concentrations $c_p/c_p^*=0$ ($\blacklozenge$), 0.1 ($\blacktriangledown$), 0.2 ($\blacklozenge$), 0.25 ($\blacktriangle$), 0.32 ($\blacksquare$), 0.4 ($\bullet$), 0.5 ($\blacklozenge$), 0.7 ($\blacktriangledown$), 0.8 ($\blacklozenge$), 1.0 ($\blacktriangle$), 1.5 ($\blacksquare$), 2.0 ($\bullet$). Lines are Mode Coupling Theory predictions for $G^\prime$ (solid lines) and $G^{\prime\prime}$ (dashed lines) (Sec.~\ref{sec:rheo_apgel}). The straight lines in (A) indicate the typical scaling in a newtonian liquid, $G^\prime \sim \omega^2$ and $G^{\prime\prime} \sim \omega$.}
\end{center}
\end{figure}

Dynamic frequency sweeps (DFS) are reported in figure~\ref{fig4:dfs_all_07} for samples below (A) and above (B) the macroscopic gelation boundary, $c_p/c_p^*=0.4$ (Fig.~\ref{fig3:macro_phase_diag}) with the frequency $\omega$ given in units of the instrument (top axis) and in units of the inverse diffusion time in the dilute limit $\tau_0=R^2/D_0\approx 4\times 10^{-2}$~s  (bottom axis).

Without polymer ($c_p/c_p^* = 0$), the response is characteristic for concentrated hard-sphere suspensions \cite{shikata94}: Viscous properties ($G^{\prime\prime}$) dominate over elastic properties ($G^\prime$), but the elasticity is still finite, i.e.~$G^\prime>0$.
 
For $c_p/c_p^* = 0.1$ and $0.2$, $G^{\prime\prime}$ still exceeds $G^\prime$ over the whole frequency range. Both mechanical moduli show the same frequency dependence, which can be described by a power-law with an exponent of about 0.55. A $G^\prime$ which is larger than for hard-spheres arises from enthalpic contributions due to the interparticle attraction. The observed  response shows interesting similarities to that measured in chemical and physical polymer gels, in particular in partially cured or weakly cross-linked materials  at the percolation point \cite{chambon86,winter86,richtering92}. This suggests that already for $0.1\lesssim c_p/c_p^* \lesssim 0.2$ a percolated network is formed. This network is transient, since, in contrast to chemical gels, the lifetime of the physical bonds between colloids is finite with bonds dynamically forming and breaking. This dynamic, transient structure is consistent with an ergodic, complete relaxation of the dynamic structure factor $f(Q,\tau)$ (Fig.~\ref{fig10:g2_dls}).

The first indication of a solid-like response is found for $c_p/c_p^*=0.25$ at large frequencies, with the crossing point of $G^\prime$ and $G^{\prime\prime}$ at $\omega\tau_0 = 10^{-2}$. This corresponds to structural relaxation times in the experimental time-window, in agreement with the fluid-like relaxation observed by dynamic light scattering (Fig.~\ref{fig10:g2_dls}). Due to the increasing strength of the attractions, the lifetime of the particle network becomes comparable to the examined time-scale (or frequency) and is long enough to cause solid-like behavior at short times, corresponding to high frequencies.

For $c_p/c_p^* = 0.32$ the frequency dependence of both, $G^\prime$ and $G^{\prime\prime}$, becomes weaker and they cross close to the low frequency limit of the investigated frequencies. This solid-like response over almost all measured frequencies indicates, upon increasing $c_p$, an increase of the structural relaxation time with an increase in the lifetime of the network and the approach to dynamical arrest. This is consistent with the drop in the long-time diffusion coefficient (Fig.~\ref{fig14:g2_dls_param2}).

When the macroscopic gelation boundary ($c_p/c_p^*=0.4$) is crossed, the frequency dependence of $G^\prime$ and $G^{\prime\prime}$ is comparable at all $c_p$, which suggests a structural relaxation time consistently larger than the experimental observation time and particle dynamics which are arrested and hence non-ergodic samples. Thus, the percolated network lacks (measurable) structural relaxation with a very long lifetime of particle bonds. Increasing $c_p$ further increases the elastic response as indicated by the increase of $G^\prime$.

Interestingly, within the gel region ($c_{p}/c_{p}^{*}\ge 0.4$), $G^{\prime\prime}$ presents a minimum at intermediate frequencies. This has already been observed for a large variety of so-called `soft glassy materials' \cite{mason95,sollich97,bibette95,cloitre00,ketz88,mackley94,koumakis_sm08} . Such a minimum suggests long-time (low frequency) structural relaxation, $\alpha$ relaxation, inside the gel phase. Its frequency could be associated with a transition from $\beta$ to $\alpha$ relaxation (both outside the experimental time window) and thus be related to the length scale over which particles diffuse before they reach the transient non-ergodicity plateau between the two processes. Since the minimum stays at $\omega R^2/D_0\approx 4\times 10^{-2}$ up to $c_{p}/c_{p}^{*}= 1$ and then shifts toward lower frequencies, this indicates a constant $\alpha$ relaxation time for gels up to $c_{p}/c_{p}^{*}= 1$ and an increasing $\alpha$ relaxation time for $c_p/c_p^*>1$.

\begin{figure}[tbp]
\begin{center}
\vspace{1cm}
\includegraphics[scale=0.4,angle=270]{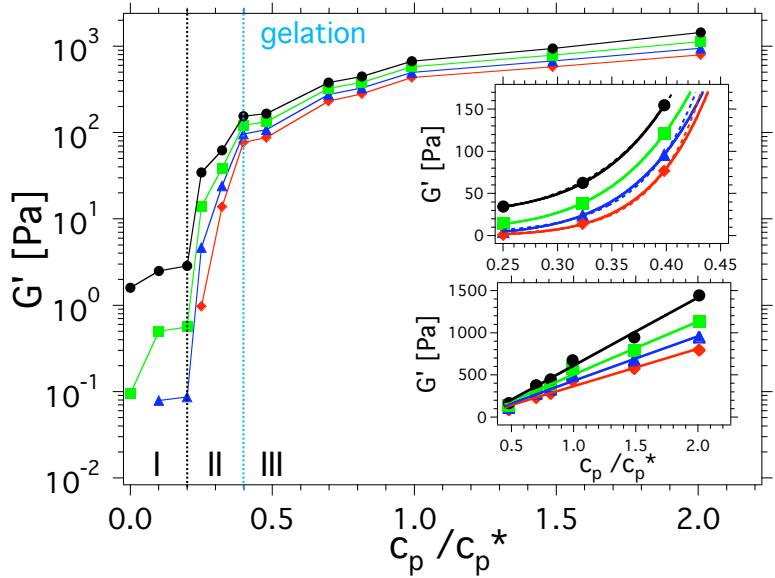}
\caption{Elastic (storage) modulus $G^\prime$ extracted from dynamic frequency sweeps as a function of polymer concentration $c_{p}/c_{p}^{*}$ at different frequencies $\omega = 0.1$ rad/s ($\blacklozenge$), 1~rad/s ($\blacktriangle$), 10~rad/s ($\blacksquare$), 100~rad/s ($\bullet$). Upper inset: $G^\prime$ as a function of $c_{p}/c_{p}^{*}$ approaching gelation ($0.25 \le c_{p}/c_{p}^{*} \le 0.4$). Solid lines represent power-law fits with exponents from about 6.2 (red) to about 8.3 (black). Dashed lines represent exponential fits. Lower inset: G$^\prime$ as a function of $c_{p}/c_{p}^{*}$ inside the gel phase ($c_{p}/c_{p}^{*} \ge 0.5$). Lines represent linear fits. }
\label{fig5:gprime_vs_cp}
\end{center}
\end{figure}

The $c_p$-dependence of the elastic modulus $G^\prime$ is summarized in figure~\ref{fig5:gprime_vs_cp}. Three different regimes can be distinguished. For $c_{p}/c_{p}^{*} < 0.25$ (region I in the figure), $G^\prime$ is very small and modestly increases with increasing $c_p$. (At the lowest frequency $\omega=0.1$~rad/s, $G^\prime$ could not be detected for $c_{p}/c_{p}^{*} < 0.25$.) For a fixed $\omega$ and increasing $c_p$, $G^\prime$ increases as a result of the increasing strength of attraction and the increasing entropic contribution from density fluctuations which are enhanced due to attraction-induced clustering. The density fluctuations will be averaged out at long times and will thus not contribute at low frequencies. Moreover, due to the low $c_p$ the bond lifetime is short and there will be no bond contribution to $G^\prime$ at long times (low frequencies). In contrast, at higher frequencies both contributions are present and $G^\prime$ thus increases with frequency.

Approaching the gelation boundary, $0.25 \le c_{p}/c_{p}^{*} \le 0.4$ (region II in the figure), $G^\prime$ shows a steep increase with increasing $c_p$ (Fig.~\ref{fig5:gprime_vs_cp}, upper inset), which reflects the strongly increasing number of permanent bonds and their increased strength which leads to a permanent, stress bearing network. The increase in $G^\prime$ can be described by a power-law or exponential dependence. The exponent of the power-law dependence increases from about 6.2 to about 8.3 with decreasing frequency. Thus, the power law dependence tends to the exponential dependence and hence the quality of the exponential fit improves with decreasing frequency. A more pronounced increase of G$^\prime$ with decreasing frequency is reminiscent of the discontinuos jump from zero to a finite shear modulus when crossing the gelation boundary as predicted by MCT \cite{dawson01}. In section \ref{sec:rheo_apgel} we compare these data to MCT predictions, which take the bond energy into account, but neglect effects of heterogeneous structure and percolation.

Within the gel region, $c_{p}/c_{p}^{*} \ge 0.5$ (region III in the figure), $G^\prime$ increases linearly (Fig.~\ref{fig5:gprime_vs_cp}, lower inset). The linear increase suggests that, once a gel is formed and saturation of permanent bonds reached, the elastic response depends on the structure of the network and the bond energy. A simple model which accounts for both contributions is proposed below (Sec.~\ref{sec:rheo_ingel}).

The dynamic frequency sweeps and the dependence of the elastic modulus $G^\prime$ on polymer concentration $c_p$ together with the dynamic light scattering results and macroscopic observations indicate the existence of two transitions: First, at $c_p/c_p^*=0.25$ network formation and the first solid-like response with a relaxation time within the experimental time window is observed. Second, at $c_p/c_p^*=0.4$ gel formation and a solid-like response with a structural relaxation time outside the experimental time window was found. While both processes imply a network structure, the dynamics of the networks, in particular the lifetime of particle bonds and hence of the whole networks, seem different.

\section{Theoretical Modeling}

\subsection{\label{sec:bondlifetime}Bond Lifetime}

Our results suggest that a crucial parameter is the time particles remain within the range of their mutual attraction, i.e.~the `bond' lifetime. This determines whether, on a given time scale, the network is transient or permanent. We estimate the bond lifetime as a function of polymer concentration $c_p$ with a simple model. It is based on the approach by Kramers to describe the escape of particles from a potential well \cite{kramers}. The first passage time of a Brownian particle within a depletion potential (Asakura-Oosawa potential) can be calculated numerically \cite{mcleish}. In order to obtain an analytical expression, we approximate the depletion potential by a ramp potential  $U(r)$ with the same depth $U_0$ ($U_0<0$) and width $\delta^*=2\xi^*R$
\begin{equation}
U(r)=\left \{ \begin{array}
{l@{\quad  \quad}l}
\infty & r \le 2R \\ U_0 \left (1-\frac{r-2R}{2\xi^* R} \right ) & 2R < r \le 2(\xi^*+1)R \\ 0 & r > 2(\xi^*+1)R
\end{array} \right.
\end{equation}
Based on the depletion potential we estimate $U_0=-\Pi_p V_{\mathrm{o}}(2R)$, where $\Pi_p$ is the osmotic pressure and $V_{\mathrm{o}}(r)$ the overlap volume of the depletion regions of two particles at distance $r$. We calculated $V_{\mathrm{o}}$ according to the generalized free volume theory (GFVT) \cite{tuinier07,lekkerkerker92,aarts02}, which accounts for the $c_p$-dependence of the polymer size and osmotic pressure (Sec.~\ref{sec:samples}). The dependence of $U_0$ on $c_p/c_{p}^{*}$ is shown in figure~\ref{fig13:tauesc} (inset). The error bars reflect the uncertainty in the size ratio $\xi^*$, which results from the uncertainty in the colloid and polymer radii. This uncertainty propagates to an uncertainty in the escape time.
The escape time $\tau_{\mathrm{esc}}$ from a ramp potential is \cite{mcleish,smith07}\\
\begin{equation}
\tau_{\mathrm{esc}} = \frac{1}{D_S^{(s)}}\int_0^{\delta^*}{\mathrm{d}}x^{\prime}e^{\beta U(x^{\prime})}\int_{-\infty}^{x^{\prime}}{\mathrm{d}}xe^{-\beta U(x)}
= \frac{\delta^{*2}}{D_S^{(s)}} \frac{e^{-\beta U_0}-(1-\beta U_0)}{(\beta U_0)^2}
\label{eq_tauesc}
\end{equation}
where $D_S^{(s)}$ is the short-time self-diffusion coefficient of a particle within the potential $U(r)$. It is estimated based on the short-time self-diffusion coefficient of a particle in a colloidal dispersion having volume fraction $\phi = 0.4$; $D_S^{(s)}\approx0.3 D_0$ \cite{pusey:leshouches}. Since the ramp potential overestimates the particle attraction in a depletion potential, $\tau_{\mathrm{esc}}$ is expected to overestimate the escape time for a depletion potential.

\begin{figure}[tbp]
\begin{center}
\vspace{1cm}
\includegraphics[scale=0.5,angle=270]{tau_esc_xfig.epsi}
\caption{\label{fig13:tauesc} Escape time $\tau_{\mathrm{esc}}$ for a particle confined to a linear ramp potential $U(r)$ as a function of polymer concentration $c_p/c_p^*$. $\tau_{esc}$ was calculated according to equation \ref{eq_tauesc}  ($\circ$) and compared to the time at which, according to rheology experiments, $G^{\prime}$ and $G^{\prime\prime}$ cross ($\blacksquare$), and to the short ($\blacklozenge$) and long ($\blacktriangle$) relaxation times obtained from light scattering at $QR=0.7$. Dashed red lines indicate the time window accessible by rheology. Inset: Dependence of $U_0$ on $c_p/c_p^*$. Line is a fit to the power- law dependence $U_0\sim (c_p/c_p^*)^{0.9}$}
\end{center}
\end{figure}

The escape time $\tau_{\mathrm{esc}}$ increases rapidly with increasing $c_{p}$ (Fig.~\ref{fig13:tauesc}). At $c_p/c_p^*\approx0.4$ it reaches lab time scales (hours) and thus indicates permanent bonds with dynamical arrest and gel formation, in agreement with the macroscopic gelation boundary (Fig.~\ref{fig3:macro_phase_diag}). For $0.1 \le c_{p}/c_{p}^{*} \le 0.32$, $\tau_{\mathrm{esc}}$ is within the experimental time window of the rheological measurements. In the rheology experiments we observe, within the accessible time window, a transition to solid-like behavior for $c_p/c_p^*=0.25$ and 0.32  (Fig.~\ref{fig4:dfs_all_07}). The frequency of the crossing point of $G^\prime$ and $G^{\prime\prime}$ (Fig.~\ref{fig13:tauesc}, filled squares) has the same order of magnitude as the calculated $\tau_{\mathrm{esc}}^{-1}$. This indicates that, approaching gelation, a particle network forms whose relaxation time, given by the lifetime of the particle bonds, determines the structural relaxation of the system. For larger $c_p$, $\tau_{\mathrm{esc}}$ and the crossing point of $G^{\prime}$ and $G^{\prime\prime}$ is beyond the time window accessible by rheology. For smaller polymer concentrations, $0\le c_{p}/c_{p}^{*} \le 0.2$, no crossing point was observed, although the calculated $\tau_{\mathrm{esc}}$ lies within the time window accessible by rheology. This suggests that in these samples the structural relaxation probed by rheology is not related to the breaking of particle bonds and the samples are rather fluids of individual particles or clusters of particles than transient network structures.

Based on the light scattering results, namely the short and long-time collective diffusion coefficients (Figs.~\ref{fig12:g2_dls_param}A, \ref{fig14:g2_dls_param2}A), we estimate the characteristic relaxation times on a length scale corresponding to the range of the potential, $\delta^{*}$, $\tau_S = \delta^{*2}/D_S^{(s)}(Q)$ and $\tau_L = \delta^{*2}/D_L^{(s)}(Q)$ and compare them to $\tau_{\mathrm{esc}}$. The ratio between the short-time self diffusion coefficient, $D_S^{(s)}$ and the collective diffusion coefficient for hard spheres with $\phi = 0.4$ as determined by dynamic light scattering is $D_S^{(s)}/D_S\approx 0.2$ \cite{pusey:leshouches}. Since attraction mainly affects the long-time decay, we use this ratio as an estimate for all $c_p$ (Fig.~\ref{fig13:tauesc}, filled diamonds). Its slight increase, almost invisible on the large vertical scale of the plot, is related to the $c_p$-dependence of the potential range $\delta^{*}$. However, this $\tau_S$ is much smaller than the calculated $\tau_{\mathrm{esc}}$. For the long-time diffusion coefficient, $D_L^{(s)} /D_L \approx 0.1$ for hard spheres  \cite{pusey:leshouches}. In this case, this relation is only valid at small $c_p$, since large $c_p$, i.e.~strong attraction, slows down the long-time self diffusion due to bonding (in analogy to repulsive and attractive glasses \cite{pham02,pham04}). This is consistent with the agreement observed at $c_{p}/c_{p}^{*} = 0$, but $\tau_L$ increasingly underestimates the time needed to diffuse a distance $\delta^{*}$, possibly being responsible for the increasingly large discrepancy between $\tau_L$ and $\tau_{\mathrm{esc}}$ when approaching the gelation boundary (Fig.~\ref{fig13:tauesc}, filled triangles). Moreover, we found that for $c_{p}/c_{p}^{*} = 0.1$ and 0.2 the rheological relaxation time is faster than the shortest time accessible in the experiments and might thus be closer to $\tau_S$ than $\tau_{\mathrm{esc}}$. We attribute this to the fact that particles or particle clusters diffuse within a shorter time than the lifetime of particle bonds. This supports the above finding that samples with $c_{p}/c_{p}^{*} \le 0.2$ are fluids of individual particles or particle clusters rather than transient or permanent networks, which start to form for $c_p/c_p^* \gtrsim 0.25$.

Finally, long-time structural relaxation has been observed by rheology in samples inside the gel region. In the gel region the bonds are so strong that they can be considered as essentially permanent (corresponding to huge $\tau_{esc}$ at $c_p/c^* > 0.4$, Fig.~\ref{fig13:tauesc}). The long-time relaxation can thus not be related to bond breaking, but might be associated with particles of different mobilities, as suggested by studies on dynamical heterogeneities \cite{puertas04}, or with different restructuring processes, such as rotation of particle groups, which then lead to the observed stress relaxation and aging.

\subsection{\label{sec:rheo_apgel}Frequency Dependence of the Moduli by Mode Coupling Theory}

Mode-coupling theory allows the shear modulus $G(t)$ to be calculated for dense suspensions and predicts the existence of a glass transition. Within this approach, the modulus is approximated by calculating the overlaps of stress fluctuations with density fluctuations in order to capture the slow structural relaxation which occurs close to the glass transition. The slow relaxation of the system is thus described by the transient density correlator. While the full mode-coupling equations possess a wavevector dependence it has been shown for the quiescent case \cite{goetze91,fuchs:prl:02} that a simplified, schematic version of the theory in which the $Q$-dependence is neglected can effectively capture the essential physics. Recent developments generalizing the theory to the case of steady-shear have shown that a similar schematic model can be used to represent the full mode coupling equations under shear, the so-called $F^{(\dot\gamma)}_{12}$ model \cite{fuchs:jcp:05,fuchs:prl:02,fuchs:faraday:03}. Within this schematic model the transient density correlator $\phi(t)$ obeys the equation of motion
\begin{equation}
\partial_t\phi(t) + \Gamma\left( \phi(t) + 
\int_0^{t}{\mathrm{d}}t' m(t-t')\left( \partial_{t'}\phi(t') +\delta\,\phi(t') \right) \right)=0 \; ,
\end{equation}
where $m(t)$ is the memory function, $\Gamma$ is the initial decay rate and $\phi(0)=1$. Introduction of the parameter $\delta$ provides an additional decay mechanism leading to long time relaxation of glassy states \cite{goetze91}. The theory assumes that $m(t)$ and the shear modulus $G(t)$ relax on the same time scale as the correlator $\phi(t)$, therefore a self-consistent approximation closing the equations of motion can be made. In the $F^{(\dot\gamma)}_{12}$ model the memory function $m(t)$ is given by
\begin{equation}
m(t) = \frac{1}{1+(\dot\gamma t)^2}(v_1\phi(t) + v_2\phi^2(t)) \; ,
\end{equation}
where $v_1$ and $v_2$ are coupling vertices chosen so as to reproduce the generic behaviour of the full, $Q$-dependent theory at the glass transition and are thus not independent, but connected by a simple algebraic relation \cite{fuchs:jcp:05,fuchs:prl:02,fuchs:faraday:03}. Typically the parameters are chosen as $v_2=2$ and $v_1=v_2(\sqrt{4/v_2}-1)+ \epsilon/(\sqrt{v_2}-1)$. In this way, both $v_1$ and $v_2$ are determined by the separation parameter $\epsilon$. The value $\epsilon=0$ corresponds to the glass transition point and positive (negative) values of $\epsilon$ correspond to statepoints in the glass (fluid). We note that for small amplitude oscillatory shear the $\dot\gamma$ dependence in $m(t)$ may be neglected. The modulus $G(t)$ is given in the $F^{(\dot\gamma)}_{12}$ model by
\begin{equation}
G(t) = v_{\sigma}( \phi^2(t) + \tilde{x}\delta ) \; ,
\end{equation}
where the modulus amplitude $v_{\sigma}$ provides an additional fit parameter. The elastic (storage) modulus $G^\prime$ and viscous (loss) modulus $G^{\prime\prime}$ are obtained by Fourier transformation
\begin{equation}
G'(\omega) + iG''(\omega) 
= i\omega\int_0^{\infty}{\mathrm{d}}t\, e^{-i\omega t}G(t)_{\dot\gamma=0}.
\end{equation}
In order to model the data, there are four free parameters: $\epsilon, \tilde{x}, v_{\sigma}$ and $\Gamma$. On a double logarithmic plot the shape of $G^\prime(\omega)$ and $G^{\prime\prime}(\omega)$ depends only on the distance from the glass transition, parameterized by $\epsilon$, and on the parameter $\tilde{x}$. The initial decay rate $\Gamma$ and amplitude $v_{\sigma}$ allow for horizontal and vertical translations, respectively. The additional decay parameter $\delta$ may then be used to fine tune $G^{\prime\prime}(\omega)$ at low frequencies.

The fundamental assumption of MCT is the description of dynamical arrest in terms of glassy dynamics, i.e.~the slowing down of particle dynamics is caused by increasing caging of particles when approaching the glass transition. This limits the range of $c_p$, i.e.~the strength of attraction, to which the model can be applied, namely the region approaching the gel transition. For low polymer concentrations, $c_{p}/c_{p}^{*}<0.2$, the dynamics are not adequately described in terms of caging. For large $c_p/c_p^*$, i.e.~inside the gel region, the model cannot reproduce the frequency dependence of $G^\prime$ and $G^{\prime\prime}$ due to the different nature of the dynamics assumed in the model (glassy) and present in the samples.

Predictions based on the $F_{12}^{(\dot{\gamma})}$ model for the region where it is assumed to be valid, namely approaching the gel transition ($c_{p}/c_{p}^{*}$ = 0.2, 0.25 and 0.32), are shown in Fig.~\ref{fig4:dfs_all_07} (lines). Theory correctly estimates the relative magnitude of $G^\prime$ and $G^{\prime\prime}$ in the high frequency limit, in particular for $c_{p}/c_{p}^{*}= 0.25$, while an increasingly larger discrepancy evolves at lower frequencies. MCT seems to associate a Newtonian fluid response to the system after structural relaxation, i.e.~at frequencies below the crossing of $G^\prime$ and $G^{\prime\prime}$, while the experimental response shows a less pronounced frequency dependence for both moduli. This could be caused by polydispersity and/or dynamical heterogeneities, which MCT does not consider but are indicated by the stretched exponential decay observed in dynamic light scattering (Sec.~\ref{sec:dls}). For $c_{p}/c_{p}^{*}$ = 0.32, $G^\prime$ agrees well with experiments, while the predicted $G^{\prime\prime}$ shows a consistently different frequency dependence. The predicted minimum appears only as a shoulder (at higher frequencies than the predicted minimum) in the experimental data. This minimum reflects the presence of an $\alpha$ relaxation in the theoretical dynamics, which is neither observed in the rheological response (Fig.~\ref{fig4:dfs_all_07}) nor in the dynamic light scattering data (Fig.~\ref{fig10:g2_dls}). We attribute this discrepancy to the above mentioned differences between the dynamics of the experimental system and the glassy dynamics implicit in the theoretical model.

\subsection{Polymer Concentration Dependence of the Elastic Modulus by Mode Coupling Theory}

\begin{figure}[tbp]
\begin{center}
\includegraphics[scale=0.5,angle=270]{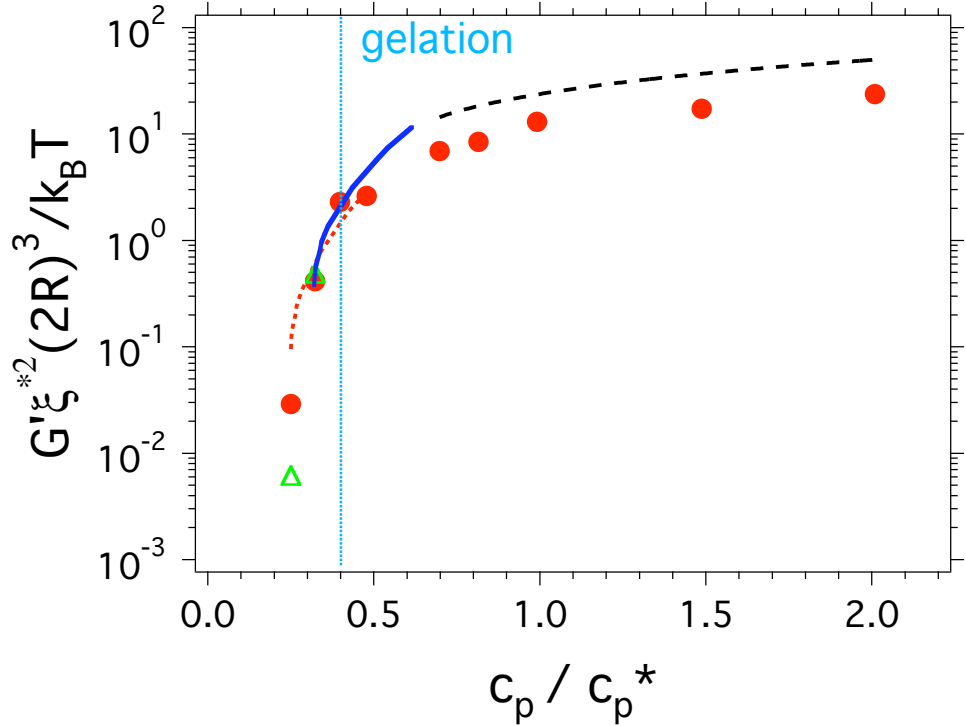}
\caption{Elastic modulus $G^\prime$ as a function of polymer concentration $c_p/c_p^*$. $G^\prime$ has been normalized by the polymer-colloid size ratio $\xi$, the particle volume $(2R)^3$ and the thermal energy ${\mathrm{k_B}}T$. The experimentally determined $G^\prime$ has been extracted from dynamic frequency sweeps at the lowest measured frequency, $\omega=0.1$~rad/s, ($\bullet$) and is compared to $F_{12}^{(\dot{\gamma})}$ ($\triangle$) and MCT-PRISM predictions for $G^\prime(\omega{\rightarrow}0)$. The dotted line is the MCT-PRISM prediction obtained for $c_p^{gel} = 0.25c_p^*$ and a scaling factor about 1/4. The solid line is is the MCT-PRISM prediction obtained for $c_p^{gel} = 0.32c_p^*$ and no scaling factor. The dashed line indicates the slope $G^\prime\sim (c_{p}/c_{p}^{*})^{0.9}$}
\label{fig7:gprime_lin_mct}
\end{center}
\end{figure}

In addition to the full frequency dependence discussed above, we now consider the $c_p$-dependence of the elastic modulus $G^\prime$ at a fixed (low) frequency $\omega$. We compare our experimental data obtained for $\omega=0.1$~rad/s (Fig.~\ref{fig7:gprime_lin_mct}, filled circles) to MCT predictions based on the $F_{12}^{(\dot{\gamma})}$ model as above (open triangles), and MCT-PRISM predictions \cite{schweizer02,zukoski03}. MCT-PRISM was recently applied to silica spheres-PS mixtures at volume fraction $\phi \approx 0.4$ and shown to correctly reproduce the $c_p$-dependence of the measured $G^\prime(\omega{=}1$~Hz) over a range of $c_p$ \cite{zukoski03}. In this study, a gel was experimentally defined on the basis of rheological measurements as a sample for which $G^\prime(\omega{=}1$~Hz$)>G^{\prime\prime}$ and $G^\prime>10$~Pa. Applied to our measurements this definition implies a gelation boundary at $c_{p}/c_{p}^{*} = 0.25$. (In contrast, according to our criterion for a gel, the sample with $c_p/c_p^*=0.25$ is not considered a gel, because it still shows structural relaxation (Fig.~\ref{fig10:g2_dls}) and thus a fluid-like response.) The range of polymer concentrations investigated in \cite{zukoski03} should thus be compared to $0.25 \le c_{p}/c_{p}^{*} \le 0.5$ in our measurements, i.e.~in the vicinity of `our' gelation boundary. For consistency with \cite{zukoski03} we also scale $G^\prime$ by $\xi^{*2}$. 

MCT-PRISM predictions for $G^\prime$ (Fig.~\ref{fig7:gprime_lin_mct}, dashed line) are considerably larger than our experimental data. In order to compare the functional dependence we rescaled the theoretical predictions. The best match with the experimentally observed trend was obtained for a scaling factor of about 1/4 (dotted line). This results in fair agreement; the increase of the experimentally observed $G^\prime$ is slightly more pronounced than predicted. A scaling factor was already proposed earlier \cite{zukoski03} to account for the difference in the structure assumed in MCT-PRISM, a homogeneous fluid, and the silica-PS gels, which show structural heterogeneities in small-angle x-ray scattering experiments \cite{shah03}. Structural heterogeneities are also present in our samples with $c_{p}/c_{p}^{*} > 0.25$ according to our static light scattering and DIC microscopy experiments (Figs.~\ref{fig8:structure_factors}, \ref{fig9:sq_lowq_zeta}). The scaling factor giving quantitative agreement between experimental data and MCT-PRISM predictions was calculated from the ratio between the particle density and the density of particle clusters (with a size corresponding to the characteristic length of structural heterogeneities) \cite{zukoski03}. The characteristic length of structural heterogeneities, derived from scattering experiments \cite{shah03}, remained constant with increasing $c_p$ for silica-PS gels. In our case, however, the characteristic length $\zeta$ shows a strong $c_p$-dependence within the range of interest $0.25 \le c_{p}/c_{p}^{*} \le 0.5$ (Fig.~\ref{fig9:sq_lowq_zeta}). These different trends of $\zeta$ lead to different functional dependencies of $G^\prime$. Nevertheless, rescaling the MCT-PRISM predictions by our $\zeta(c_p)$ makes the agreement between MCT-PRISM predictions and our experimental data worse.
 
There is some ambiguity in the determination of the gelation boundary. If we choose $c_p/c_p^*=0.32$ instead of 0.25, this would possibly agree better with the definition of the gelation boundary by the theory, namely the transition from zero to a finite value of $G$. Moreover, this shift might be justified by differences in the structure factor between the silica-PS mixtures and our samples. With a gelation boundary at $c_{p}/c_p^*=0.32$, no scaling factor is needed (Fig.~\ref{fig7:gprime_lin_mct}, solid line). For $0.32< c_{p}/c_{p}^*<0.5$ the MCT-PRISM predictions describe the data well, while there are considerable discrepancies for $c_{p}/c_{p}^*>0.5$ and no data for larger $c_{p}/c_{p}^*$ are available for comparison \cite{zukoski03}. They might be due to the difficulty to treat non-equlibrium states well inside the gel region or to account for the changes in attraction range and strength beyond the overlap concentration $c_p^*$.


These comparisons indicate that MCT and MCT-PRISM do not entirely capture the mechanisms responsible for the evolution of the shear moduli approaching and entering the gel region. Structural heterogeneities migh cause the observed discrepancies in both cases: First, structural heterogeneities can induce dynamical heterogeneities, which are not included in the glassy dynamics of the $F_{12}^{(\dot{\gamma})}$ model. Second, the MCT-PRISM model is based on structure factors of fluid-like equilibrium structures, not heterogeneous structures.

\subsection{\label{sec:rheo_ingel}Polymer Concentration Dependence of the Elastic Modulus Inside the Gel Region}

The structure of a gel can be considered as closely packed fractal clusters \cite{shih90,krall98}. We estimate the effect of changes in this microscopic structure on the $c_p$-dependence of the elastic properties. 
For large clusters, the elastic behavior of the gel will be dominated by the deformation of clusters (strong-link regime) and the elastic constant of a cluster, $K_{\zeta}$, is expected to depend on the size of its backbone, i.e.~decreases with increasing cluster size. For small clusters, intercluster links will deform before clusters deform (weak-link regime). The number of particle-particle links between clusters is smaller than the average number of particle bonds inside a cluster, as evidenced by confocal microscopy measurements of similar colloidal gels \cite{smith07,smith04}. Hence the elastic constant of the system will be dominated by the elastic constant of intercluster links, $K_l$. In both cases, the total elastic contant scales as $G\sim K/\zeta$ \cite{shih90,marangoni99,wu01}.

Due to the large colloid volume fraction $\phi \approx 0.4$, our gels consist of small clusters (Fig.~\ref{fig9:sq_lowq_zeta}) and hence are expected to be in the weak-link regime. To obtain a scaling relation between $G^\prime$ and $c_p$, we have to determine the $c_p$-dependence of $K_l$ and $\zeta$. The correlation length $\zeta$ sharply decreases just above the gelation boundary but then remains approximately constant well inside the gel region (Fig.~\ref{fig9:sq_lowq_zeta}). In a minimal model, the elastic constant of intercluster links, $K_l$, is expected to depend on the number $m$ of particle-particle contacts between clusters and the interaction between two particles at contact, $U_0$, i.e.~$K_l \sim mU_0$. We assume that $m$ does not depend on $c_p$. The $c_p$-dependence of $U_0$ (Fig.~\ref{fig13:tauesc}, inset) can be fitted by a power-law dependence, $U_0\sim (c_p/c_{p}^{*})^{0.9}$, inside the gel region. This results in $G^\prime \sim K_l/\zeta \sim mU_0/\zeta \sim U_0 \sim (c_p/c_p^*)^{0.9}$. This scaling is in agreement with $G^\prime(c_p)$ observed in experiments (Fig.~\ref{fig7:gprime_lin_mct}, straight solid line). Therefore, this simple model seems to capture the essential mechanism leading to the elasticity of the gels, namely the intercluster links.

\section{Conclusions}

We investigated the structural, dynamical and rheological properties of colloid-polymer mixtures with an intermediate colloid volume fraction, $\phi = 0.4$, as a function of increasing polymer concentration, corresponding to increasing interparticle attraction. These samples covered a broad range from liquids to gels. The macroscopic gelation boundary was determined by tube inversion.

The structure of the samples was investigated by static light scattering and microscopy. Within the liquid we observed the formation of increasingly larger structures, especially when approaching the gelation boundary. Increasing attraction induces the formation of particle clusters which, at sufficiently large attraction, interconnect to form a space-spanning network. The maximum cluster size and maximum structural heterogeneity is observed at the gelation boundary. Within the gel region, increasing attraction leads to a more uniform structure with a reduction in the characteristic length scale. This trend is reminiscent of critical behavior expected for phase separation that is arrested by gelation.

Within the entire liquid phase, the dynamics shows an ergodic response. Upon approaching gelation, the short-time (in-bond) diffusion as well as the long-time diffusion, which leads to the final structural relaxation of the system, slows down. The more pronounced slowing down of the long-time decay indicates the approach of gelation and its increasingly stretched exponential form suggests that heterogeneities in the density fluctuations are increased and the distribution of length scales is broadened, reminiscent of clustering. This is consistent with the increasing correlation length observed in static light scattering and microscopy. This suggests that structural and dynamical heterogeneities, namely clustering and the formation of transient networks, are precursors of gel formation.

Rheological measurements in the linear viscoelastic regime show, with increasing polymer concentration, a shifting of the crossing point of the elastic and viscous moduli corresponding to a transition from a liquid-like to a gel-like response at a characteristic frequency (time). This characteristic time could be related to the `bond' lifetime estimated by the time needed to escape from the interparticle attraction. Bond breaking was found to be the dominating process close to the gelation boundary, while at lower polymer concentrations the relaxation appears to be related to particle or cluster diffusion.

The elastic modulus of samples approaching the gelation boundary have been compared to MCT predictions within the $F_{12}^{(\dot{\gamma})}$ model \cite{fuchs:faraday:03}. The predictions reproduce the dependence of the modulus on attraction strength, i.e.~polymer concentration. However, the frequency-dependence of the modulus shows discrepancies. This could be due to structural and dynamical heterogeneities, which are not included in the theory.

We compared the same data, the dependence of the elastic modulus on polymer concentration, also to MCT-PRISM predictions \cite{zukoski03}. Comparison between experiment and theory requires a consistent definition of the gelation boundary and/or a scaling factor. This is necessary, because, again, heterogeneities are present, which are not considered by the theory. With an appropriate choice, agreement can be obtained in a limited range of polymer concentrations, i.e.~attraction strengths.

In the gel region, the shear moduli show a solid-like behavior, with $G^{\prime}$ weakly frequency-dependent and always larger than $G^{\prime\prime}$, which shows a minimum in the frequency dependence. At fixed frequency, $G^{\prime}$ increases almost linearly with polymer concentration. We suggest that this is consistent with a fractal model for gel elasticity in the so-called weak-link regime (similar to low volume fraction gels \cite{krall98,trappe00,shih90}). Within this model, the almost linear dependence of $G^\prime$ results from the increase of the energy of inter-cluster links with increasing polymer concentration, while, at the same time, the cluster size in the gel remains approximately constant. This implies that, due to the heterogeneous structure of the samples, the elasticity of the gels is dominated by cluster-cluster links rather than particle-particle bonds.

\section{Acknowledgements}

This work was funded by the Deutsche Forschungsgemeinschaft (DFG) within the German-Dutch Collaborative Research Centre Sonderforschungsbereich-Transregio 6 (SFB-TR6), Project Section A6. The D{\"u}sseldorf-Crete collaboration was supported by the EU Network of Excellence `SoftComp'. M. Laurati would like to acknowledge E. Zaccarelli for stimulating discussions.

\bibliography{proceeding_george_v9}
\newpage

\end{document}